\documentclass[]{spie}  

\usepackage{amsmath,amsfonts,amssymb}
\usepackage{graphicx}
\usepackage{comment} 
\usepackage{multirow} 
\usepackage{multicol} 
\usepackage[colorlinks=true, allcolors=blue]{hyperref}

\title{Learning to segment prostate cancer by aggressiveness from scribbles in bi-parametric MRI}

\author{Audrey Duran}
\author{Gaspard Dussert}
\author{Carole Lartizien}
\affil{Univ Lyon, CNRS, Inserm, INSA Lyon, UCBL, CREATIS, UMR5220, U1206, F‐69621, Villeurbanne, France}

\authorinfo{Further author information:\\Carole Lartizien: E-mail: carole.lartizien@creatis.insa-lyon.fr, Telephone: +33 (0)4 72 43 81 48}

\begin{document} 
\maketitle

\begin{abstract}
In this work, we propose a deep U-Net based model to tackle the challenging task of prostate cancer segmentation by aggressiveness in MRI based on weak scribble annotations. This model extends the size constraint loss proposed by Kervadec et al. \cite{kervadec_constrained-cnn_2019} in the context of multiclass detection and segmentation task. This model is of high clinical interest as it allows training on prostate biopsy samples and avoids time-consuming full annotation process. Performance is assessed on a private dataset (219 patients) where the full ground truth is available as well as on the ProstateX-2 challenge database, where only biopsy results at different localisations serve as reference. 
We show that we can approach the fully-supervised baseline in grading the lesions by using only 6.35\% of voxels for training. We report a lesion-wise Cohen's kappa score of $0.29 \pm 0.07$ for the weak model versus $0.32 \pm 0.05$ for the baseline. We also report a kappa score ($0.276 \pm 0.037$) on the ProstateX-2 challenge dataset with our weak U-Net trained on a combination of
ProstateX-2 and our dataset, which is the highest reported value on this challenge dataset for a segmentation task to our knowledge.
\end{abstract}

\keywords{Prostate cancer, Weakly-supervised learning, Semantic segmentation, Multiparametric MRI, Deep Learning, Gleason Score, CNN constraints}

\section{INTRODUCTION}
\label{sec:intro} 

Deep learning has become the state-of-the-art approach for the processing and analysis of many medical imaging problems, including detection and segmentation tasks \cite{tajbakhsh_embracing_2020}. However, the vast majority of the proposed models are fully supervised, meaning that they require a large amount of annotated data to perform well, which is hard to acquire due to the high expertise and time needed for the annotation. To alleviate the need for full annotation, weakly-supervised approaches have been  widely studied over the past few years \cite{chan_comprehensive_2021}. They consist in learning from partial labels, such as image tags,
 bounding boxes or scribbles. 

In this paper, we address the challenging task of segmentation and characterisation of prostate cancer (PCa) aggressiveness in bi-parametric MRI (bp-MRI) based on scribbles. 
This task has only been barely addressed and only in a fully supervised manner, for PCa detection \cite{cao_joint_2019} or segmentation by aggressiveness \cite{de_vente_deep_2020, duran2022papier}. We recently tackled this issue based on a U-Net backbone architecture combining multi-task learning and an attention model \cite{duran2022papier}.
Besides the time gain with weak labels, this problem is of high clinical interest as cancer aggressiveness, graded by Gleason score (GS), is mostly determined based on biopsy samples, that characterize the lesion at a specific location but do not allow to know the extent of the lesion. This kind of approach could also allow the inclusion of the ProstateX-2 dataset, which is the only public database, so far, containing PCa patient MRI exams, each associated to the centroid coordinates and the aggressiveness of its reported lesions \cite{prostatex2}.
\section{MATERIALS AND METHODS}

\subsection{Data description}
\label{sec:data}

Two datasets are used in this study, both containing lesion's aggressiveness information (by Gleason score):
\begin{itemize}
    \item Our private dataset. The main dataset used in this study consists of a series of axial T2 weighted (T2w) and apparent diffusion coefficient (ADC) MR images from 219 patients, acquired in clinical practice at our partner clinical center. All patients underwent a radical prostatectomy and their exams were fully annotated (including lesions with their associated Gleason score and prostate contours) at pixel-level based on prostatectomy gold standard ground truth. Imaging was performed on three different scanners from different constructors and magnetic field strength : 67 exams on a 1.5T Symphony scanner (Siemens Medical Systems), 126 on a 3T scanner Discovery scanner (General Electric) and 26 on a 3T Ingenia scanner (Philips Healthcare).
    \item ProstateX-2 train set. ProstateX-2 challenge dataset~\cite{prostatex2} is composed of a training set of 99 patients with 112 lesions. These data were acquired on 3T MAGNETOM Trio and Skyra (Siemens Medical Systems) scanners with different imaging parameters. The ground truth consists of the coordinates of each lesion's center based on biopsy results, with its associated Gleason score. Note that the ProstateX-2 test set was not used since no ground truth is available and the challenge webpage is closed.
\end{itemize}

Lesions distribution according to the GS group is detailed in Table \ref{tab:lesionsclass} for both datasets, where GS 3+3 and GS $\ge$ 8 represent the less and most aggressive cancers, respectively.
Input MRI exams are series of axial T2 weighted (T2w) and apparent diffusion coefficient (ADC), stacked in two channels.\\

\begin{table}[!t]
\caption{Lesions distribution by Gleason score (GS) in both datasets.} \label{tab:lesionsclass}
\centering
\begin{tabular}{cccccc}
\hline
&  \textbf{ GS 3+3} & \textbf{GS 3+4} &\textbf{ GS 4+3}  & \textbf{GS $\ge$ 8} & \textbf{Total}\\ 
Our dataset& 104 & 126 & 56 & 52 & 338 \\
ProstateX-2& 36 & 41 & 20 & 15 & 112 \\
\hline
\end{tabular}
\end{table}

Multichannel 3D input T2w and ADC images of both datasets were resampled to a $1\times1$ mm$^2$  
pixel size
and automatically cropped to a $96\times96$ pixels region on the image's center. The slice thickness was preserved and is around 3 for our private dataset as well as for most of ProstateX-2 patients. Most of our dataset acquisitions contain 24 slices while it is variable but around 20 slices for ProstateX-2 challenge dataset. Intensity was linearly normalized into [0, 1] by patient and channel.

Circular scribble annotations were automatically generated on both our private and ProstateX-2 datasets, as follows:
\begin{itemize}
    \item Our dataset : circular scribbles with radius $\le 4$ pixels (i.e. 4 mm) were drawn by randomly sampling, on each transverse slice of the the training dataset, one location in the prostate zone as well as one location in each reported lesion.
    If no scribble with a 4 pixels radius could fit in the lesion, radius was reduced until the scribble could meet the overlapping criterion.
    \item ProstateX-2 : circular scribbles with radius of 4 pixels (i.e. 4 mm) were drawn on each slice of the training dataset. Lesion scribbles were centered at the reported ground truth lesion center. As prostate annotations are not available, scribbles in the prostate gland were inferred from the lesion localisation as follows~: 
    the $x$ coordinate of the prostate scribble was chosen at 11 mm from the lesion's center in direction to the center of the image. For the $y$ position, it was defined at 11 mm in direction of the center of the image if the lesion was in the transition zone (TZ) and kept unchanged if the lesion was in the peripheral zone (PZ). Scribbles drawn in the slice of the lesion coordinates were reported to the 2 adjacent slices, and only annotated slices were used for training. Consistency of the obtained scribbles was checked visually.
\end{itemize}
Table \ref{tab:scribbles_percent} reports the ratio between the number of annotated voxels in the scribbles and in the full segmentation masks for each class, corresponding to 6.35\% in total. Note that even if the ratio seems high for some classes (32.77\% for GS6 for example), it still constitutes a huge time saving as the radiologist only needs to pick one point for each lesion and the circle scribble would be generated automatically around this center with the chosen radius.

\begin{table}[]
\caption{Ratio between the number of annotated voxels in the scribbles and in the full segmentation masks for each class in our private dataset.}\label{tab:scribbles_percent}
\centering
\begin{tabular}{ccccccc}
\hline
\textbf{Class}   & \textbf{Prostate} & \textbf{GS 6 } & \textbf{GS 3+4 }& \textbf{GS 4+3} & \textbf{GS $\ge$ 8} & \textbf{Total} \\ 
Ratio (\%) & 4.92     & 32.77 & 29.36  & 25.85  & 18.36    & 6.35 \\ \hline
\end{tabular}
\end{table}

\subsection{A weak multi-class segmentation model of PCa in bp-MRI}

\subsubsection{Constrained loss function for weak supervision based on Kervadec et al. \cite{kervadec_constrained-cnn_2019}.}
Kervadec et al. recently proposed a new loss function for partially annotated data. It combines a partial cross-entropy (CE) term $\mathcal{H}$ estimated on the annotated voxels and a constraint term $\mathcal{C}$, that penalizes predicted segmentation whose size is outside a defined interval $[a, b]$ :

\begin{equation}
  \mathcal{H}(S)+ \lambda \mathcal{C}(V_S)
  \label{eq: weak_loss}
\end{equation}
where $\lambda$ is a positive constant weighting the two terms, $V_S=\sum_{p \in \Omega}S_p$ with  $S_p$ the softmax probability at pixel $p$ in the image domain $\Omega$, and the functions $\mathcal{H}$, $\mathcal{C}$ are given by:

\begin{multicols}{2}
\begin{equation}
   \mathcal{H}(S)=- \frac{1}{\vert\Omega_{A}\vert}\sum_{p \in \Omega_{A}} \log \left(S_{p}\right) 
\end{equation}
\begin{equation}
    \mathcal{C}(V_S)= \left\{
    \begin{array}{lll}
        (V_S-a)^2, &  \mbox{if }V_S < a \\
        (V_S-b)^2, & \mbox{if } V_S > b \\
        0, & \mbox{otherwise}
    \end{array}
\right.
\label{eq: bounds}
\end{equation}
\end{multicols}
with $\Omega_{A}$ the domain of annotated pixels, $a$ and $b$ the lower and upper bounds of the region of interest respectively.\\
In this study, we address a multi-class problem with class labels $c$ ranging from 1 to $c$=6 for the background, the overall prostate area, GS 6, GS 3+4, GS 4+3 and GS~$\ge8$ lesions, respectively.
To adapt to a multi-class output with $C$ classes and account for class imbalance, the global loss term is obtained by a weighted sum of the loss terms (1) applied independently on each class $c \ne 1$ as follows : 
\begin{equation}
    \sum_{c \in \{2,...,6\}} w_c(\mathcal{H}(S_c)+ \lambda \mathcal{C}(V_{S_c})),
\end{equation}
with $w_c$ the weight for class $c$. 

We used this constraint at the image level with \textit{image-tag priors}, following the definition in Kervadec et al. \cite{kervadec_constrained-cnn_2019}, that is enforcing the presence of the target class by setting $a=1$ and $b=\vert\Omega\vert$ (the image domain) or the absence of the target with parameters $a=b=0$. When the target is present in the image, the predicted object will always be smaller than the upper bound $b$ but the lower bound $a$ remains crucial. In addition, both bounds are used when the target is absent of the image, making this constraint non-trivial.

\subsubsection{Backbone architecture}

The model used in this work is based on a standard four blocks U-Net \cite{ronneberger_u-net:_2015}, with batch normalization layers to reduce over-fitting and leaky ReLU activations. 
It produces a 6-channels segmentation maps, corresponding class labels $c$ ranging from 1 to 6 for the background, the overall prostate area, GS 6, GS 3+4, GS 4+3 and GS~$\ge8$ lesions. This standard U-Net architecture was shown efficient for the detection and grading of PCa with bp-MRI \cite{duran2022papier}.

\subsection{Experiments}

In this study, we perform different experiments:
\begin{itemize}
    \item we first evaluate performance of our proposed scribble based weak U-Net based on the image tags bounds constrained loss
    by comparing it to the fully supervised approach based on our private fully annotated dataset ;
    \item we then include the weakly annotated data from ProstateX-2 challenge and train the weak segmentation model on both ProstateX-2 and our datasets.
\end{itemize}

\paragraph{Implementation details}
Each model was trained and validated using a 5-fold cross-validation, with 4 replicates for each cross-validation experiment.
Patients were distributed in the folds so as to balance as much as possible the number of lesions per class, and the number of patients from each database (ProstateX-2 and ours) for the hybrid training of the weakly supervised architecture. Training and evaluation were conducted on the whole patient 3D volumes.
Data augmentation was applied during the training phase to reduce overfitting. Final lesion maps were estimated from the labeled maps outputted from the U-Net decoder branch using a 3-connectivity rule to identify the connected components. Lesions smaller than 26 voxels (ie. 78 mm$^3$) were removed. 
All networks were trained using Adam and a L2 weight regularization with $\gamma= 10^{-4}$. The initial learning rate was set to $10^{-3}$ with a 0.5 decay after 25 epochs without validation loss improvement. Batch size was set to 32. Hyperparameters were tuned with random grid search. We set $\lambda$ value for the constraint in Eq. (\ref{eq: weak_loss}) to $10^{-5}$, $w_c = 0.22$ for the cancer classes, $w_c = 0.12$ for the prostate ($c=2$) (these values were found to be the optimal to account for class imbalance in our dataset). The fully supervised baseline was trained with the sum of the crossentropy and Dice weighted losses. The pipeline was implemented in python with the Keras-Tensorflow 2.4 library.

\subsection{Performance evaluation}

Performance evaluation was conducted from the final lesion maps generated by each model 
using quadratic weighted Cohen's kappa coefficient (as proposed in the ProstateX-2 challenge) for both datasets and free-response receiver operating characteristics (FROC) curve for our private dataset. Evaluation on our dataset was performed as follows~: if a predicted lesion intersected a ground truth lesion with at least 10\% of its volume, it was counted as a true positive. Then, to compute the FROC, we measured the percentage of detected lesion (sensitivity) as a function of the mean number of false positive lesion detections per patient at different probability thresholds. 
The Cohen kappa score was computed based on the 4-classes confusion matrix encompassing GS 6, GS 3+4, GS 4+3 and GS $\ge 8$ lesion categories. 

For ProstateX-2 dataset, as lesion contours are not available, we followed the evaluation method proposed by De Vente et al. \cite{de_vente_deep_2020} : for each reference lesion center (whose coordinates and GS are provided by the ProstateX-2 challenge organizers), the lesion was assigned the Gleason score corresponding to the cluster containing it, if any. If the reference lesion center did not intersect any detected lesion, it was reported as a GS 6 lesion. 

\section{RESULTS}

Table \ref{tab:results} reports segmentation performance of all considered models, namely the fully supervised U-Net model (referred to as 'fully supervised') trained with a weighted cross-entropy + Dice loss, a weakly supervised U-Net trained with partial cross-entropy only (referred to as 'Partial CE') and the proposed weakly supervised U-Net trained with a loss term combining partial cross-entropy and image tags priors (referred to as 'Partial CE + Tags'). 

Performance achieved for these three models trained and tested on our private dataset indicate, as expected, that
the highest values of kappa coefficient of $0.324 \pm 0.053$ and sensitivity at 2FP of 0.649 $\pm 0.033$ are obtained with the fully supervised U-Net. However, performance of the proposed weak U-Net model with partial CE + Tags is close to the fully supervised model with kappa coefficient of $0.289 \pm 0.072$ and sensitivity at 2FP of 0.587 $\pm 0.053$. According to a Wilcoxon signed-rank test, the difference is not significant with a p-value of 0.165 for the kappa metric and 0.409 for the sensitivity. The poor performance achieved with the weak model trained on partial CE only (kappa coefficient of $-0.054 \pm 0.193$ and sensitivity at 2FP of 0.016 $\pm 0.008$) confirms the positive impact of the \textit{image-tags} constraint.

The highest reported performance on the ProstateX-2 dataset is achieved with the proposed weak model trained on both datasets with the combined partial CE and image tags priors. Note that the reported kappa score of $0.276\pm0.037$ is, as far as we know, the highest value reported in the literature so far.
The proposed constraint loss term, however, is not sufficient to enable correct training on the ProstateX-2 data only, as revealed by the very low kappa metric of  $-0.002 \pm 0.003$.
This can be explained by the small size of the 
ProstateX-2 train dataset (99 patients), for which only slices with a lesion were included. In addition, prostate scribbles in the normal prostate tissue for this dataset were generated from the known lesion's positions, as described above, 
thus inducing less variability than among our private dataset which contains prostate scribbles annotations in lesion-free prostate slices. 

\begin{figure*}[h!]
\centering
\caption{\label{fig:predictions}Prediction for several images from validation sets. The images from the first three rows come from our dataset while images from the two last rows come from the ProstateX-2 challenge dataset. Second column: fully supervised U-Net trained with our dataset. Third and fourth columns: weakly supervised U-Net trained on scribble annotations of our dataset and of both ProstateX-2 and our datasets, respectively.}
\includegraphics[width=0.9\textwidth]{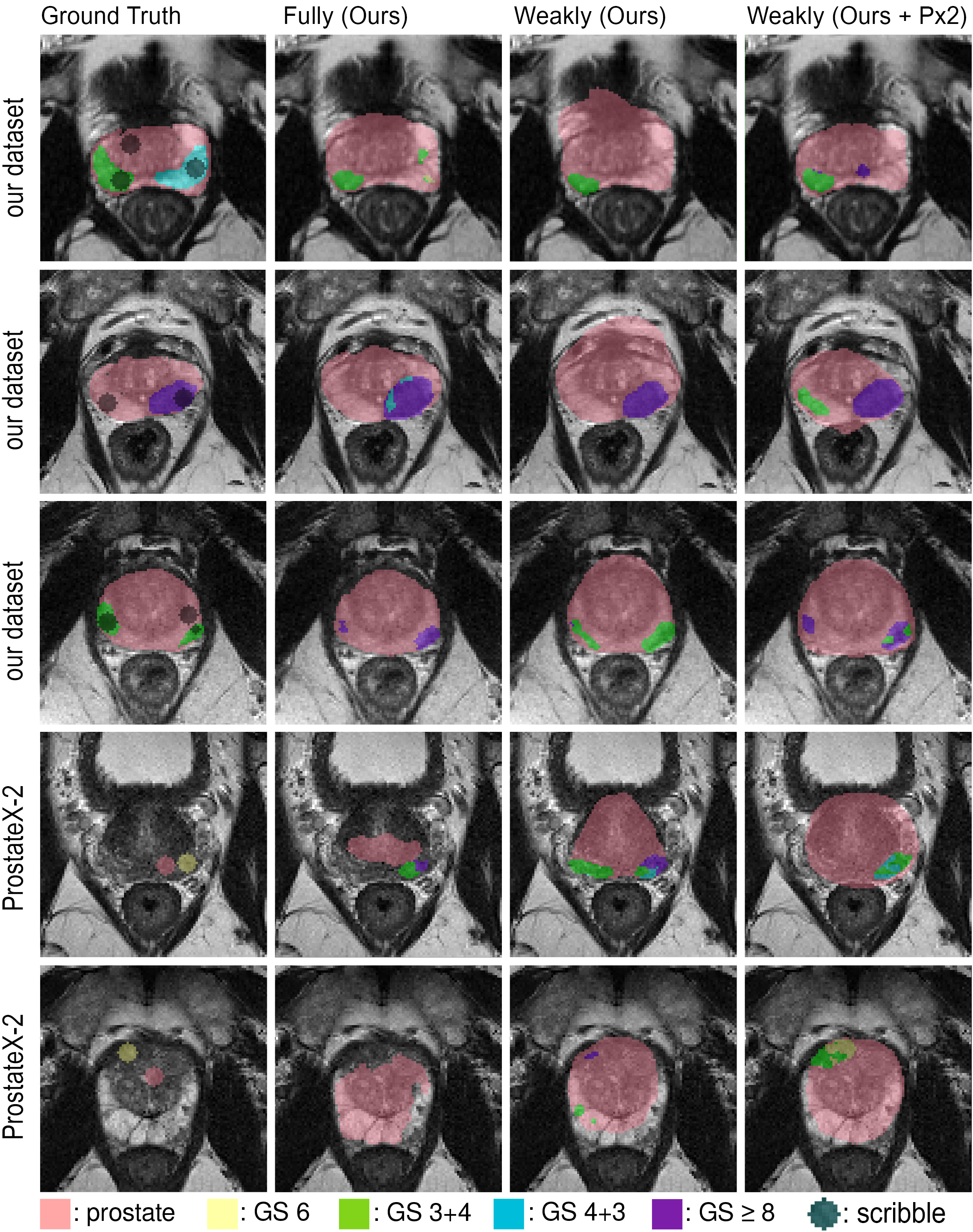} 
\end{figure*}

Concerning visual results (see Figure \ref{fig:predictions}), we can see that even with weak annotations, segmentation maps are of good quality. Segmentation of the prostate is slightly less precise on our dataset with the weak model than with the supervised one, when trained on our dataset only, but segmentation accuracy improves when trained on both ProstateX-2 and our datasets. One interesting result concerns generalization on ProstateX-2 dataset~: visually, the worst performing model is the supervised one. We can see that even the prostate tissue is not well segmented. It seems that the fully-supervised model is more prone to overfitting and is likely not to consider the prostate shape but independent pixels unlike the weakly supervised models.

\begin{table}[]
\resizebox{\textwidth}{!}{ 
\begin{tabular}{|l|l|cccc|} 
\hline
\multicolumn{1}{|c|}{\multirow{2}{*}{Method}} & \multicolumn{1}{c|}{\multirow{2}{*}{\begin{tabular}[c]{@{}c@{}}Training\\  dataset\end{tabular}}} & \multicolumn{3}{c|}{Performance on our dataset}                                                    & \multicolumn{1}{c|}{Performance on Px2} \\ \cline{3-6} 
\multicolumn{1}{|c|}{}                        & \multicolumn{1}{c|}{}                                                                             & \multicolumn{1}{c|}{Kappa} & \multicolumn{1}{c|}{Sensi at 2FP} & \multicolumn{1}{c|}{Dice prostate} & \multicolumn{1}{c|}{Kappa}              \\ \hline\hline
Fully supervised & \multicolumn{1}{l|}{ours}                                                                         & \multicolumn{1}{c|}{$\mathbf{0.324 \pm 0.053}$}     & \multicolumn{1}{c|}{$\mathbf{0.649 \pm 0.033}$}             & \multicolumn{1}{c|}{$0.799 \pm 0.004$}              & \multicolumn{1}{c|}{$0.013 \pm 0.082$}                   \\ \hline
Partial CE & \multicolumn{1}{l|}{ours}                                                                         & \multicolumn{1}{c|}{$-0.054\pm0.193$}      & \multicolumn{1}{c|}{$0.016 \pm 0.008$}             & \multicolumn{1}{c|}{$0.081 \pm 0.006$}              & \multicolumn{1}{c|}{$-0.005\pm0.019$}                   \\ \hline
Partial CE + Tags & \multicolumn{1}{l|}{ours}                                                                         & \multicolumn{1}{c|}{$0.289 \pm 0.072$}      & \multicolumn{1}{c|}{$0.618 \pm 0.044$}             & \multicolumn{1}{c|}{$0.800 \pm 0.017$}              & \multicolumn{1}{c|}{$0.047 \pm 0.060$}                   \\ \hline
Partial CE + Tags & \multicolumn{1}{l|}{Px2}                                                                         & \multicolumn{1}{c|}{$0.134 \pm 0.144$}      & \multicolumn{1}{c|}{$0.026 \pm 0.012$}             & \multicolumn{1}{c|}{$0.121 \pm 0.010$}              & \multicolumn{1}{c|}{$-0.002 \pm 0.003$}                   \\ \hline
Partial CE + Tags & ours + Px2 & \multicolumn{1}{c|}{$0.262 \pm 0.061$}      & \multicolumn{1}{c|}{$0.587 \pm 0.053$ }             & \multicolumn{1}{c|}{$\mathbf{0.802 \pm 0.005}$}              & \multicolumn{1}{c|}{$\mathbf{0.276\pm0.037}$}                   \\ \hline             
\end{tabular}}
\label{tab:results}
\caption{Segmentation performance. 
The results correspond to the average metrics obtained on 4 replicates of 5-fold cross-validation. CE : cross-entropy. Tags : Image tags bounds constraint.}
\end{table}

\section{CONCLUSIONS}

Results reveal that we can approach fully-supervised performance by training a U-Net model with only 6.35\% of pixels annotation thanks to a loss term combining partial cross-entropy and a size constraint loss derived from that of Kervadec et al. \cite{kervadec_constrained-cnn_2019}. These results are particularly interesting considering the time gain when using scribbles instead of pixel-level annotation. In addition, scribble ground truth is very relevant for PCa screening as most of the databases rely on biopsy results.
Such scribble-based segmentation models are likely to ease the inclusion of data from different sources (e.g. different centers or scanners) to tackle the recurrent domain shift problem in medical image segmentation.

In Kervadec et al. \cite{kervadec_constrained-cnn_2019}, the authors also proposed a higher supervision with adapted bounds for the size constraints, defined from the dataset statistics. This approach seemed very interesting for the lesion segmentation task as the higher the Gleason score, the bigger the lesion. However, preliminary results did not show any improvement when using more precise size constraints based on the observed lesion size distribution in comparison to the rough image tag. The size constraints are difficult to define for this problem, as some small elements of a lesion might be present on a slice but not representing the global 3D lesion size and more than one object can be present on the same slice, on the contrary to segmentation tasks considered in Kervadec et al.

Perspectives would be to investigate hybrid training on both fully and weakly annotated datasets and lesion shape constraints tailored to our application.

\appendix    

\acknowledgments 
 
This work was supported by the RHU PERFUSE (ANR-17-RHUS-0006) of Université Claude Bernard Lyon 1 (UCBL), within the program “Investissements d'Avenir” operated by the French National Research Agency (ANR). It was performed within the framework of the LABEX PRIMES (ANR-11-LABX-0063) of Université de Lyon operated by the French National Research Agency (ANR).

\bibliography{refs} 

\begin{thebibliography}{1}

\bibitem{kervadec_constrained-cnn_2019}
Kervadec, H., Dolz, J., Tang, M., Granger, E., Boykov, Y., and Ben~Ayed, I.,
  ``Constrained-{CNN} losses for weakly supervised segmentation,'' {\em Medical
  Image Analysis}~{\bf 54},  88--99 (May 2019).

\bibitem{tajbakhsh_embracing_2020}
Tajbakhsh, N., Jeyaseelan, L., Li, Q., Chiang, J.~N., Wu, Z., and Ding, X.,
  ``Embracing imperfect datasets: {A} review of deep learning solutions for
  medical image segmentation,'' {\em Medical Image Analysis}~{\bf 63},  101693
  (July 2020).

\bibitem{chan_comprehensive_2021}
Chan, L., Hosseini, M.~S., and Plataniotis, K.~N., ``A {Comprehensive}
  {Analysis} of {Weakly}-{Supervised} {Semantic} {Segmentation} in {Different}
  {Image} {Domains},'' {\em Int J Comput Vis}~{\bf 129},  361--384 (Feb. 2021).

\bibitem{cao_joint_2019}
Cao, R., Mohammadian~Bajgiran, A., Afshari~Mirak, S., Shakeri, S., Zhong, X.,
  Enzmann, D., Raman, S., and Sung, K., ``Joint {Prostate} {Cancer} {Detection}
  and {Gleason} {Score} {Prediction} in mp-{MRI} via {FocalNet},'' {\em IEEE
  Transactions on Medical Imaging}  (2019).

\bibitem{de_vente_deep_2020}
De~Vente, C., Vos, P., Hosseinzadeh, M., Pluim, J., and Veta, M., ``Deep
  {Learning} {Regression} for {Prostate} {Cancer} {Detection} and {Grading} in
  {Bi}-parametric {MRI},'' {\em IEEE Trans. Biomed. Eng.} ,  1--1 (2020).

\bibitem{duran2022papier}
Duran, A., Dussert, G., Rouvi{\`e}re, O., Jaouen, T., Jodoin, P.-M., and
  Lartizien, C., ``{ProstAttention}-{Net}: {A} deep attention model for
  prostate cancer segmentation by aggressiveness in {MRI} scans,'' {\em Medical
  Image Analysis}~{\bf 77},  102347 (Apr. 2022).

\bibitem{prostatex2}
Litjens, G., Debats, O., Barentsz, J., Karssemeijer, N., and Huisman, H.,
  ``Prostatex challenge data.'' The Cancer Imaging Archive (2017).

\bibitem{ronneberger_u-net:_2015}
Ronneberger, O., Fischer, P., and Brox, T., ``U-{Net}: {Convolutional}
  {Networks} for {Biomedical} {Image} {Segmentation},'' {\em arXiv:1505.04597
  [cs]}  (2015).

\end{thebibliography}
\bibliographystyle{spiebib} 

\end{document}